\documentclass[12pt]{iopart}
\usepackage{iopams}  
\usepackage{graphicx}
\usepackage{epstopdf}
\usepackage{cite}
\begin{document}

\title[SPM of graphene moir\'e structures]{Multichannel scanning probe microscopy and spectroscopy of graphene moir\'e structures}

\author{Yu S Dedkov$^1$ and E N Voloshina$^2$}

\address{$^1$SPECS Surface Nano Analysis GmbH, Voltastra\ss e 5, 13355 Berlin, Germany}
\address{$^2$Humboldt-Universit\"at zu Berlin, Institut f\"ur Chemie, 12489 Berlin, Germany }
\ead{Yuriy.Dedkov@specs.com}
\ead{Elena.Voloshina@hu-berlin.de}

\begin{abstract}
The graphene moir\'e structures on metals, as they demonstrate both long (moir\'e) and short (atomic) scale ordered structures, are the ideal systems for the application of scanning probe methods. Here we present the complex studies of the graphene/Ir(111) system by means of 3D scanning tunnelling and atomic force microscopy/spectroscopy as well as Kelvin-probe force microscopy. All results clearly demonstrate a variation of the moir\'e and atomic scale contrasts as a function of the bias voltage as well as the distance between the scanning probe and the sample, allowing to discriminate between topographic and electronic contributions in the imaging of a graphene layer on metals. The presented results are accompanied by the state-of-the-art density functional theory calculations demonstrating the excellent agreement between theoretical and experimental data. 
\end{abstract}




\maketitle



\section{Introduction}\label{introduction}

Graphene, a two-dimensional layer of carbon atoms arranged in a honeycomb lattice, is in a focus of many surface science researchers since its fascinating transport properties were demonstrated~\cite{Novoselov:2005,Zhang:2005,CastroNeto:2009}. These experiments renewed the interest to a graphene layer (or monolayer of graphite in a former time) on transition metal surfaces~\cite{Tontegode:1991ts,Wintterlin:2009,Dedkov:2012book,Batzill:2012}. This area of the surface science can be traced back in the middle of 60s when the spectroscopic studies of the catalytic properties of transition metal surfaces were started. At that time the carbon layer segregated at high temperature at the surface of the transition metals was considered as a poison inert layer which blocks any catalytic activity. However, at the present time, a growth of graphene on metals, via segregation from the metal bulk or decomposition of hydrocarbons at the transition metal surface, are considered as the most prominent methods for the preparation of large (up to several tens of inches) high-quality graphene sheets~\cite{Bae:2010}. Further, these graphene layers can be transferred on the insulating or polymer support and then scaled down for the preparation of tiny high-speed transistors, gas sensors, touch screens, etc~\cite{Li:2009,Bae:2010,Tao:2012if}.

From the other, more fundamental point of view, the interaction at the graphene/metal interface is interesting problem by itself~\cite{Wintterlin:2009,Khomyakov:2009,Voloshina:2012c}. Here one can point out the importance of this topic as the resistance of the graphene/metal contact in any electron- or spin-transport devices will be determined by the structural as well as electronic properties of such interface. With this respect it is worth to mention that the full understanding of the bonding mechanism and the corresponding electronic properties of the graphene/metal interface is far from the full understanding~\cite{Voloshina:2012c}. 

As is widely accepted now, the graphene/metal systems can be subdivided on two big classes of the ``strongly'' and ``weakly'' bonded graphene on the metallic surface. Here, for the ``strongly'' bonded graphene layer [e.\,g., graphene/Ni, graphene/Co, graphene/Rh] the significant modification of the graphene-derived electronic states is observed and the Dirac cone is not preserved in such systems, opposite to the case of the ``weakly'' bonded graphene on metals [e.\,g., graphene/Cu, graphene/Ir, graphene/Pt, graphene/Au]. Such situations are found for both lattice-matched (graphene/$3d$-metal interfaces) and lattice-mismatched systems (graphene/$4d$-metal and graphene/$5d$-metal interfaces). It is interesting to note that the later systems are proposed to be used as templates for the preparation of ordered arrays of well-defined metal clusters, which can be used in future storage and catalytic applications.

Presently, for the investigation of the electronic structure of the graphene/metal interfaces, the spectroscopic methods are widely used: photoelectron spectroscopy, PES, gives and information about doping level of a graphene layer in the system (x-ray PES, XPS) or about the behaviour of the graphene-derived states in the valence band (angle-resolved PES, ARPES)~\cite{Voloshina:2011NJP,Starodub:2011a,Dedkov:2012book} and Raman spectroscopy gives an information about doping level as well as about the defect concentration in the graphene layer~\cite{Malard:2009,Entani:2011gr,Starodub:2011a}. These methods have limited space resolution and are not sensitive to the variation of the electronic structure of graphene on the nm-scale. However, the local information about electronic structure of graphene on the scale of the moir\'e structure or atomic scale can be obtained only via application of the scanning probe methods [scanning tunnelling microscopy/spectroscopy (STM/STS) and/or non-contact atomic force microscopy (NC-AFM)]. The recently developed 3D non-contact atomic force microscopy (3D NC-AFM) with conductive scanning tip~\cite{Albers:2009ig} and Kelvin-probe force microscopy (KPFM) methods with atomic scale resolution~\cite{Melitz:2011,Mohn:2012gh} can give such information combining the structural and electronic structure information on the sub-nm scale. Such attempts were recently undertaken for graphene/Rh(111)~\cite{Voloshina:2012a}, graphene/Ir(111)~\cite{Voloshina:2013dq,Boneschanscher:2012bg}, and BN/Rh(111)~\cite{Koch:2012by} (BN is an isostructural to graphene single layer consisting of B and N atoms), where the combination of scanning probe microscopy and spectroscopy was used. These studies demonstrate the selective interaction between the different sites of the moir\'e structure and the scanning tip and the inversion of the imaging contrast of the moir\'e structure as well as the similar effects on the atomic scale as a function of the distance between sample and the tip (metallic and CO-terminated tips were used). However, the more comprehensive studies are still required.

This manuscript presents the complex study of the graphene moir\'e structure on Ir(111). We combine STM, 3D NC-AFM, and KPFM methods in order to get information about local structural and electronic properties of the graphene/metal system on the nm- and atomic scale. The simultaneous application of these methods allows to get the spatial distribution of the tunnelling current as well as of the interaction forces between sample and the scanning tip. Our results demonstrate the clear variation of the imaging contrast during scanning probe measurements as a function of the distance between sample and scanning tip as well as a bias voltage, that allows to separate between topographic and electronic contributions in all imaging modes (STM and NC-AFM). We also found a significant variation of the local electrostatic potential in the graphene moir\'e on Ir(111) that is reflected in the contrast changes in the $I(z)$ maps and in the KPFM picture. All results are accompanied by the state-of-the-art density functional theory (DFT) calculations and good agreement between experimental and theoretical data is found.

\begin{figure}
\center
\includegraphics[scale=0.3]{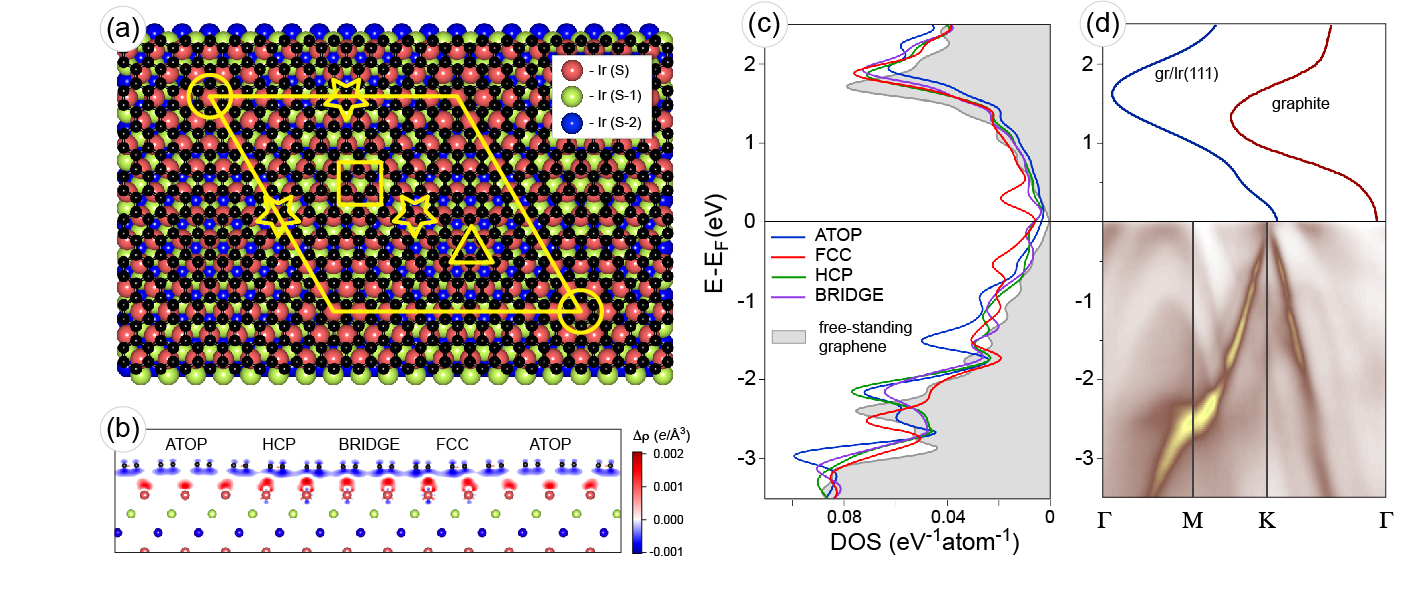}
\caption{(a) Crystallographic structure of the graphene/Ir(111) systems (small black and large coloured spheres are carbon and Ir atoms, respectively). Circles, squares, triangles, and stars denote ATOP, HCP, FCC, and BRIDGE positions, respectively. (b) Side view of the graphene/Ir(111) interface with the corresponding difference electron density, $\Delta\rho (r)=\rho_{gr/Ir(111)}(r)-\rho_{Ir(111)}(r)-\rho_{gr}(r)$, plotted in units of $\mathrm{e/\AA}^3$. (c,d) Distribution of the density of the electronic states in the vicinity of $E_F$ for graphene/Ir(111). In (c) the carbon-atom projected site-resolved DOS is shown for different high-symmetry adsorption places of carbon atoms on Ir(111). The corresponding DOS for free-standing graphene calculated for the supercell of the similar size is shown as shaded area. The corresponding C $K$ $1s\rightarrow\pi^*$ NEXAFS and ARPES data (along $\Gamma-\mathrm{M}-\mathrm{K}-\Gamma$ direction of the graphene-derived Brillouin zone) for the graphene/Ir(111) system are combined in (d) for the same energy range. The NEXAFS spectrum for graphite is shown for comparison.}
\label{structure}
\end{figure} 

\section{Experimental and computational details}\label{experimental}

\textit{Experiment.} The STM and NC-AFM measurements were performed in two different modes: constant current (CC) or constant frequency shift (CFS) and constant height (CH). In the first case the topography of sample, $z(x, y)$, is studied with the corresponding signal, tunnelling current ($I_T$) or frequency shift ($\Delta f$) of an oscillating sensor, used as an input for the feedback loop. In the later case the $z$-coordinate of the scanning tip is fixed with a feedback loop switched off, that leads to the variation of the distance $d$ between tip and sample. In such experiments $I_T(x, y)$ and $\Delta f(x, y)$ maps are measured for different $z$-coordinates. 

In a more sophisticated 3D NC-AFM measurements the $I_T(z)$ and $\Delta f(z)$ curves were collected on a grid. Then these curves were combined in one data set, which was used for the analysis~\cite{Albers:2009ig,Baykara:2010a}. As a reference ``0''-point for the $z$-scale the simultaneously collected STM data set (picture) was used allowing also for the possible drift and creep post-correction of the piezo drive (additionally to the atom-tracking mode used for the drift compensation). The large $z$-scale (up to 5\,nm above the STM contact point) $I_T(z)$ and $\Delta f(z)$ curves were measured in the beginning and in the end of every data set accounting for the thermal drift of the frequency shift of the sensor.

All STM/NC-AFM/KPFM images were collected at room temperature with SPM Aarhus 150 equipped with KolibriSensor\texttrademark\ from SPECS~\cite{Torbruegge:2010cf,Voloshina:2012a,Voloshina:2013dq} with Nanonis Control system. In all measurements the sharp W-tip was used which was cleaned \textit{in situ} via Ar$^+$ sputtering. In presented STM images the tunnelling bias voltage, $U_T$, is referenced to the sample and the tunnelling current, $I_T$, is collected by the tip, which is virtually grounded. During the AFM measurements the sensor was oscillating with the resonance frequency of $f_0=999161$\,Hz and the quality factor of $Q=45249$. The oscillation amplitude was set to $A=100$\,pm or $A=300$\,pm (specified in the text for every discussed experiment). Separate connection of the current and oscillation outputs allows to avoid any cross-talk between STM and AFM signals.\\
\textit{Sample preparation.} The graphene/Ir(111) system was prepared in ultrahigh vacuum station for STM/NC-AFM studies according to the recipe described in details in Refs.~\cite{Coraux:2008,Coraux:2009,Voloshina:2013dq} via cracking of ethylene: $T=1100^\circ$\,C, $p=1\times10^{-7}$\,mbar, $t=15$\,min. This procedure leads to the single-domain graphene layer on Ir(111) of very high quality that was verified by means of low-energy electron diffraction (LEED) and STM. Prior to this procedure the Ir(111) substrate was cleaned via series of sputtering/flash-annealing cycles. The base vacuum was better than $7\times10^{-11}$\,mbar during all experiments.\\
\textit{Calculations.} The crystallographic model of graphene/Ir(111) presented in Fig.~\ref{structure}(a) was used in DFT calculations, which were carried out using the projector augmented plane wave method~\cite{Blochl:1994}, a plane wave basis set with a maximum kinetic energy of 400\,eV and the PBE exchange-correlation potential~\cite{Perdew:1996}, as implemented in the VASP program~\cite{Kresse:1994,Kresse:1996,Kresse:1996a}. The long-range van der Waals interactions were accounted for by means of a semiempirical DFT-D2 approach proposed by Grimme~\cite{Grimme:2006}. The studied system is modelled using supercell, which has a $(9\times9)$ lateral periodicity and contains one layer of $(10\times10)$ graphene on a four-layer slab of metal atoms. Metallic slab replicas are separated by ca. $20$\,\AA\ in the surface normal direction. To avoid interactions between periodic images of the slab, a dipole correction is applied~\cite{Neugebauer:1992}. The surface Brillouin zone is sampled with a $(3\times3\times1)$ $k$-point mesh centred the $\Gamma$ point.

\begin{figure}
\center
\includegraphics[scale=0.55]{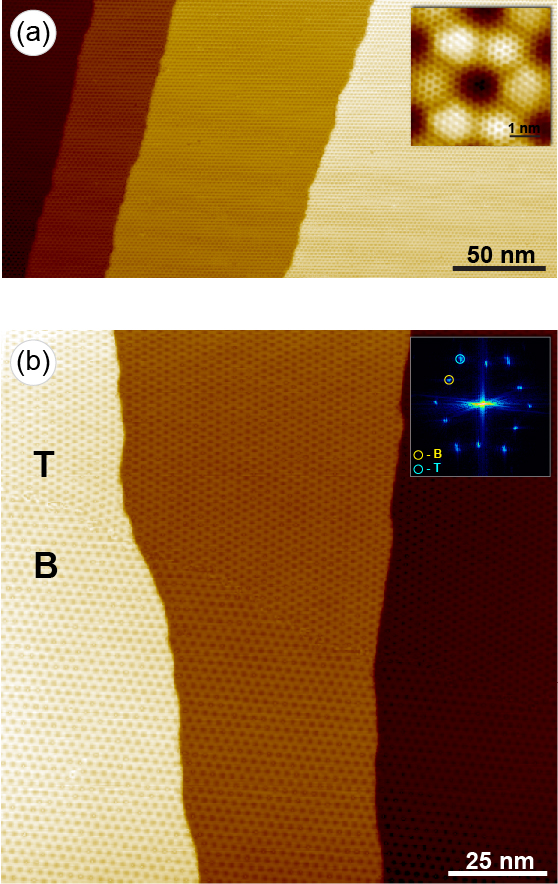}
\caption{(a) Large scale STM image of graphene/Ir(111) showing single domain graphene layer ($300\times150$\,nm$^2$, $U_T=0.5$\,V, $I_T=1.9$\,nA) (Image was cutted from the originally collected $300\times300$\,nm$^2$ scan frame). The inset shows a zoomed image of the single graphene moir\'e ring on Ir(111) ($4.55\times4.55$\,nm$^2$, $U_T=0.5$\,V, $I_T=10$\,nA). (b) STM image of two slightly rotated graphene domains (T and B, respectively) on Ir(111). The inset shows the corresponding FFT image with the respective hexagonal spots marked by the coloured circles.}
\label{STM_LargeScale}
\end{figure} 

\section{Results and discussion}\label{results}

\subsection{STM and bias dependence}\label{STMandBIAS}

The top view of the crystallographic structure of the graphene/Ir(111) system is shown in Fig.~\ref{structure}(a). Here graphene has a periodicity of 10 carbon rings on top of 9 Ir(111) unit cells that leads to the size of the moir\'e structure of $24.6$\,\AA. This structure was used in the calculations of the optimised crystallographic arrangement, which result is shown in Fig.~\ref{structure}(b) with the distribution of the electron density in this system as well as for the calculation of the density of the valence band states [Fig.~\ref{structure}(c)]. In the previous studies (see Refs.~\cite{Pletikosic:2009,Busse:2011}) it was suggested that graphene on Ir(111) behaves like a free-standing without strong modification of the graphene-derived valence band states around the Fermi level ($E_F$). However, graphene layer in this system is slightly buckled. The distances between a graphene layer and Ir(111) for the high symmetry positions are: $3.58$\,\AA\ for ATOP, $3.28$\,\AA\ for FCC, $3.27$\,\AA\ for HCP, and $3.315$\,\AA\ for BRIDGE, respectively. Also there is a slight $p$-doping of graphene (electron charge transfer from graphene to Ir) [Fig.~\ref{structure}(b)], that leads to the shift of the Dirac point above $E_F$ by $\approx 100$\,meV. This value of doping is in very good agreement with previously published value of $100\pm20$\,meV derived from ARPES data for graphene/Ir(111) (see Ref.~\cite{Pletikosic:2009} and present data in Fig.~\ref{structure}(d)). Variation of the distance between graphene and Ir(111) surface in the moir\'e unit cell leads to the modulation of the interaction strength between carbon and Ir atoms. It can be reflected in the different carbon-atom-projected density of states around $E_F$ for different hight-symmetry arrangement positions of the graphene/Ir(111) system [Fig.~\ref{structure}(c)]. One can clearly see that, compared to the free-standing graphene (density of states is shown as a shaded area), for graphene/Ir(111) several additional peaks appeared in the energy range close to $E_F$. [Here we would like to emphasise that calculations for the free-standing graphene and for the graphene/Ir(111) interface were carried out in the supercell geometry of the  $(10\times10)$ periodicity in order to perform the correct comparison of the density of states in both systems.] For example, for the $ATOP$ position two peaks arise at $E-E_F=-0.95$\,eV and at $E-E_F=-1.49$\,eV and for the $FCC$ place two peaks at $E-E_F=-0.53$\,eV and at $E-E_F=+0.32$\,eV. These peaks can be identified in the experimentally obtained electronic structure of graphene/Ir(111), in ARPES and NEXAFS data [Fig.~\ref{structure}(d)]. Generally, the C $K$-edge NEXAFS spectrum of the graphene-based system consists of two regions reflecting the $1s\rightarrow\pi^*$ and $1s\rightarrow\sigma^*$ transitions~\cite{Wessely:2005,Wessely:2006}. Comparison of C $K$ $1s\rightarrow\pi^*$ NEXAFS spectra of graphite and graphene/Ir(111) is shown in Fig.~\ref{structure}(d), upper panel for energies above $E_F$, where the theoretically predicted feature at $E-E_F=+0.32$\,eV can be attributed to the shoulder at $E-E_F\approx+0.45$\,eV. From the comparison of the calculated DOS below $E_F$ and ARPES data we can conclude that peaks in DOS can be assigned to the mini Van Hove singularities appearing due to the opening of the mini-gaps as a result of crossing of the main graphene $\pi$-band with replica bands. These replica bands appear as a result of the moir\'e superlattice of the graphene/Ir(111) system and due to the disturbing the periodic potential (due to interaction of graphene and Ir valence band states), which modulates the carbon-lattice potential.

Fig.~\ref{STM_LargeScale}(a) shows the representative $300\times150$\,nm$^2$ STM image of a graphene layer on Ir(111) prepared at $1100^\circ$\,C. One can clearly see that in this case the single domain graphene was grown with extremely small number of defects. The inset of this figure shows a single moir\'e ring of this graphene layer. STM experiments on graphene/Ir(111)  usually performed at the low bias voltages and they demonstrate the so called \textit{inverted contrast} with respect to the expected variation of the hight in the moir\'e unit cell of this system: the topographically highest ATOP places are imaged as dark in STM images and the lowest FCC and HCP places are imaged as bright areas [inset of Fig.~\ref{STM_LargeScale}(a)]. This fact immediately indicates the importance of the graphene-Ir interaction (hybridisation and charge transfer) in this system, although a graphene layer here is always described as nearly free-standing~\cite{Pletikosic:2009,Busse:2011,Voloshina:2013dq}.

\begin{figure}
\center
\includegraphics[scale=0.6]{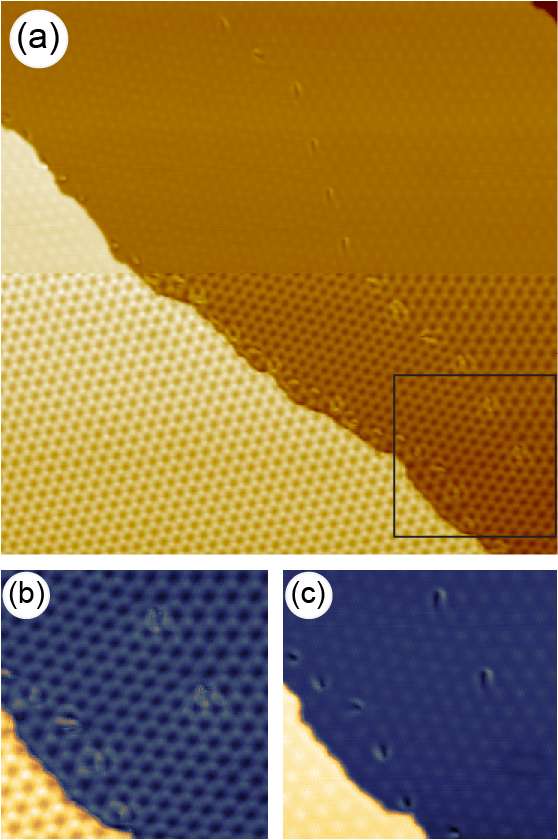}
\caption{(a) Demonstration of the contrast inversion ``on-the-fly'' during STM imaging of the graphene moir\'e structure on Ir(111): bottom and upper parts are imaged at $U_T=-0.3$\,V and $U_T=-1.6$\,V, respectively. Scan size: $100\times100$\,nm$^2$. (b) and (c) are atomically resolved STM images of the zoomed region [marked by the solid rectangle in (a)] obtained at $U_T=-0.3$\,V and $U_T=-1.6$\,V, correspondingly. Scan size: $30\times30$\,nm$^2$.}
\label{STM_RotDomains}
\end{figure}

A slight reduction of the graphene's synthesis temperature leads to the little misorientation in the graphene layer and the appearing of the rotational domains [Fig.~\ref{STM_LargeScale}(b) and Fig.~\ref{STM_RotDomains}]. Fig.~\ref{STM_LargeScale}(b) shows the STM image of the area of the graphene/Ir(111) system prepared at $1000^\circ$\,C. Two domains with the moir\'e structures rotated by $30^\circ$ with respect to each other are present: the corresponding Fast-Fourier-Transformation (FFT) image is shown as an inset. As was shown in Ref.~\cite{NDiaye:2008qq}, in this case the misorientation of the graphene rings is of only $3^\circ$ in the neighbouring domains. 

The interesting result for the slightly misoriented graphene domains on Ir(111) is shown in Fig.~\ref{STM_RotDomains}. Here the large scale image in (a) shows the STM results obtained during changing ``on-the-fly'' of the tunnelling bias voltage: bottom part was collected at $U_T=-0.3$\,V and the upper part was acquired at $U_T=-1.6$\,V. The corresponding small scale images in (b) and (c) are moir\'e as well as atomically resolved images of the region marked by solid rectangle in (a). Here one can clearly resolve three graphene domains, which are oriented by $0^\circ$, $15^\circ$, and $30^\circ$ with respect to each other. In this case the graphene layers are misoriented by $1.5^\circ$ and $3^\circ$, respectively. These results clearly demonstrate that the imaging contrast in CC STM of graphene/Ir(111) depends on the bias voltage (sign and value) between tip and the sample surface, indicating the existence of the graphene-metal interaction in this system. According to the comparison of the site-resolved carbon projected DOS of grapehene/Ir(111) and free-standing graphene discussed earlier, the inversion of the imaging contrast in STM can be unambiguously attributed to the appearance of the mini Van Hove singularities in the electronic structure of the graphene/Ir(111) system. The DOS peak at $E-E_F=-0.53$\,eV corresponding to the FCC places is responsible for the brighter imaging of these places in CC STM, whereas ATOP places become brighter in CC STM at larger bias voltages, where two states at $E-E_F=-0.95$\,eV and $E-E_F=-1.49$\,eV start to contribute in the tunnelling process. This effect of the contrast inversion is not expected at the positive bias voltages as supported by the experiment~\cite{Voloshina:2013dq}. It is interesting to note that inversion of the imaging contrast in CC STM with changing of the bias voltage is observed simultaneously for different graphene domains indicating the local character of interaction between graphene and Ir(111) that defines the imaging contrast.

Generally, the existence of the mini Van Hove singularities due to the moir\'e structure cannot explain the observation of the inversion of the imaging contrast in CC STM of graphene on Ir(111). The following consideration might give some hints for the explanation of this effect. Considering the electronic structure of graphene/Ir(111) in the vicinity of the $\mathrm{K}$ point of the Brillouin zone one can imaging 6 replica cones surrounding the main cone of the $\pi$ band. Electronic bands avoid-crossing mechanism leads to the appearance of the mini gaps where these additional cones cross the main cone with the increased density of states at particular wave vectors, leading to the appearance of the mini Van Hove singularities in the density of states distribution. At the same time the Ir $5d$ orbitals, which have projections on the $z$-axis (perpendicular to the graphene plane), i.\,e. $d_{xz}$, $d_{yz}$, and $d_{z^2}$ are in the trigonal planar crystal field of the $D_{3h}$ symmetry due to the graphene layer on top, that leads to the corresponding splitting of these states and ordering them in the same energy sequence~\cite{Zuckerman:1965}. As we found, the interaction between a graphene layer and Ir(111) is not negligible. Therefore we can expect that the contribution of the corresponding Ir $5d$-states in the density of states of the respective mini Van Hove singularity will depend on the high symmetry adsorption position of the carbon atoms on Ir(111): $d_{xz}$ and $d_{yz}$ orbitals will contribute in the states corresponding to the ATOP position at $E-E_F=-0.95$\,eV and $E-E_F=-1.49$\,eV and $d_{z^2}$ will contribute in the states corresponding to the FCC position at $E-E_F=-0.53$\,eV. This consideration explains the corresponding contrast inversion that might be connected with the appearance of the mini Van Hove singularities and their interaction with Ir $5d$-states of the respective symmetry.

\begin{figure}
\center
\includegraphics[scale=0.5]{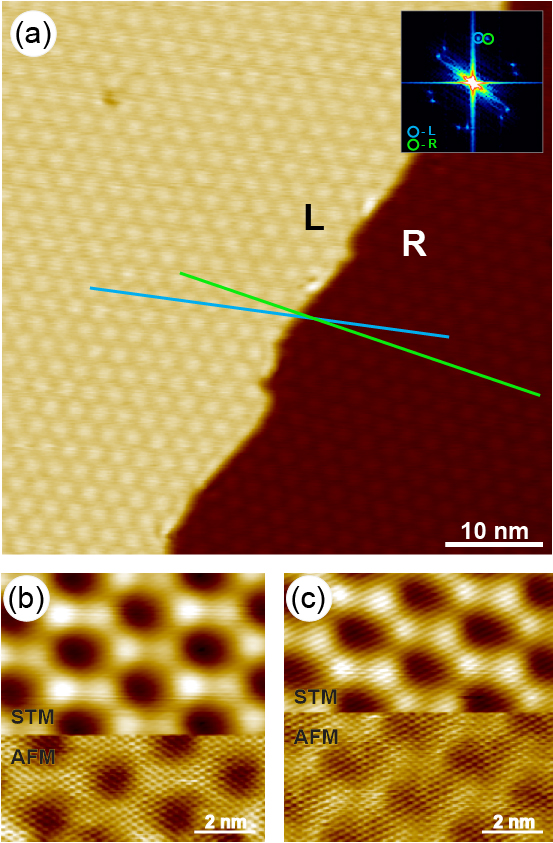}
\caption{(a) CFS NC-AFM image of the graphene/Ir(111) system. Imaging parameters: scan size $56\times56$\,nm$^2$, $U_T=-0.23V$, set point $\Delta f=-0.7$\,Hz, $A=300$\,pm. (b,c) Switching between CC STM and CFS AFM ``on-the-fly'' (the corresponding areas are marked as STM and AFM). Imaging parameters: scan size $10\times10$\,nm$^2$, $I_T=7.6$\,nA, $\Delta f=-0.675$\,Hz, $A=300$\,pm. Bias voltages are $U_T=+0.46$\,V and $U_T=-0.2$\,V for (b) and (c), respectively.}
\label{AFM_RotDomains}
\end{figure} 

\subsection{NC-AFM and contrast inversion between STM and AFM}\label{CFSAFM}

The graphene/Ir(111) system was imaged in the CFS NC-AFM mode (Fig.~\ref{AFM_RotDomains}). These images were acquired in the attractive regime (see below) and the ATOP positions are imaged as bright areas with a faint trace of the atomic resolution. Similarly to the previously discussed STM data, here we also imaged slightly rotated, with respect to each other, graphene domains [marked as ``L'' and ``R'' in Fig.~\ref{AFM_RotDomains}(a)]. Here the angle between the corresponding directions of the graphene moir\'e structure is $15^\circ$ corresponding to the misalignment of the graphene rings of only $1.5^\circ$. 

\begin{figure}
\center
\includegraphics[scale=0.45]{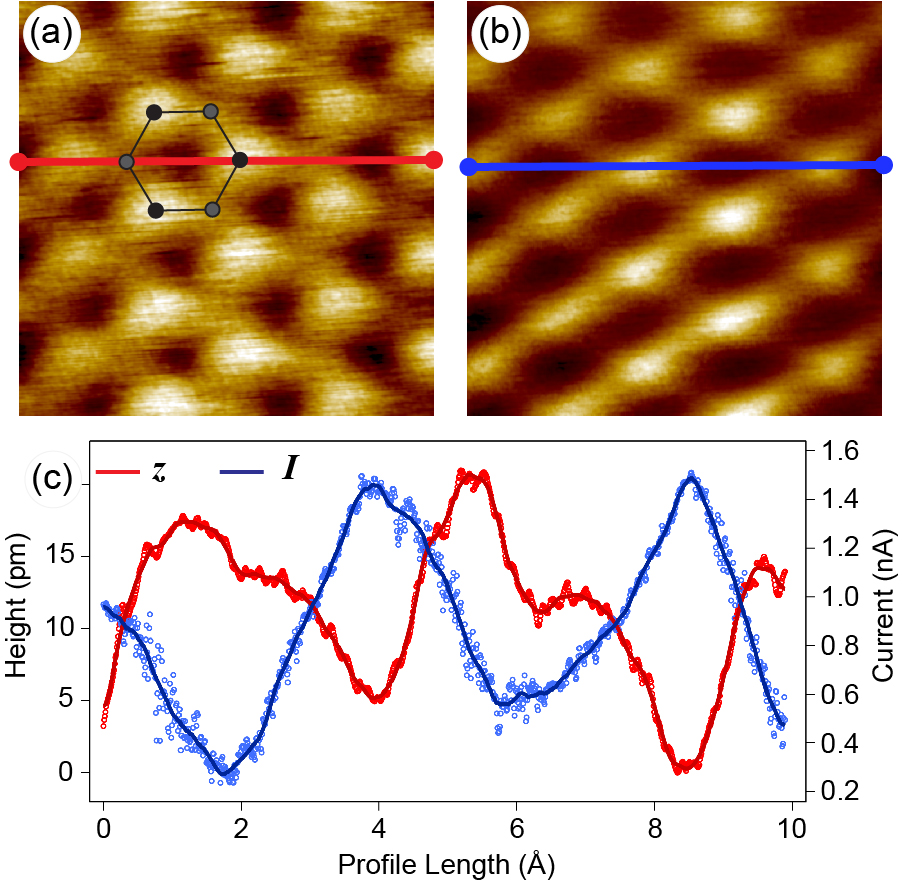}
\caption{CFS NC-AFM images of the ATOP position of the graphene/Ir(111) system: (a) topography $z$-channel and (b) the simultaneously recorded tunnelling current, $I_T$-channel. Imaging parameters: scan size $10\times 10$\,\AA$^2$, $\Delta f=-0.81$\,Hz, $A=300$\,pm, $U_T=+0.01$\,V. (c) Corresponding line profiles for both channels.}
\label{AFM_AtRes}
\end{figure} 

One of these domains (left, ``L'') was used for the more systematic studies and for the comparison between STM and AFM imaging of a graphene moir\'e structure on Ir(111). We imaged this area in the regime where we could successfully switch ``on-the-fly'' between CC STM and CFS AFM during scanning that allows us to trace the changes in the imaging contrast between two modes [Fig.~\ref{AFM_RotDomains}(b,c)]. In both modes the moir\'e structure as well as atomic structure are clearly resolved (depending on the bias voltage in STM mode the corrugation of the moir\'e structure can smear the atomic contrast). For the tunnelling bias voltages used here, graphene/Ir(111) is imaged in the \textit{inverted} contrast during CC STM (see discussion in Sec.~\ref{STMandBIAS}). One can clearly see that compared to CC STM, in the CFS NC-AFM the same area is imaged now in the \textit{true} contrast (which is inverted with respect to those in CC STM). Now the topographically more high ATOP places are imaged as bright areas and the imaging contrast between HCP and FCC places depends on the bias voltage applied between tip and sample. Generally, the interaction force between tip and sample is a sum of van der Waals (attractive, $F\propto -z^{-2}$) and electrostatic interactions ($F\propto -z^{-1}\propto -{U_T}^2$) between macroscopic tip and sample and the local chemical forces. Therefore, the change of the tunnelling bias can lead to the variation of the local contrast between regions of the nearly similar hight but with different electrostatic potential distribution [Fig.~\ref{AFM_RotDomains}(b,c)].

Successful CFS AFM imaging of graphene/Ir(111) gives us a chance to perform an atomically resolved imaging with the aim to identify every carbon atom in the graphene unit cell. It is a well know imaging problem in STM that for graphite only every second carbon atom, which has no neighbour in the adjacent layer, can be imaged~\cite{Tomanek:1987a,Hembacher:2003a}. Only CH AFM imaging performed in the repulsive regime (where Pauli repulsion plays a dominant role) allowed recently to image every carbon atom in the unit cell of graphite. In our case of a graphene layer on Ir(111), due to the relatively weak interaction between a graphene layer and metallic substrate, one can expect a successful imaging of the carbon ring. The results of this experiment are compiled in Fig.~\ref{AFM_AtRes}. Here the ATOP place of the graphene/Ir(111) system was measured in CFS NC-AFM attractive regime mode and simultaneously topography ($z$-channel) [Fig.~\ref{AFM_AtRes}(a)] and tunnelling current ($I_T$-channel) [Fig.~\ref{AFM_AtRes}(b)] were acquired. The corresponding ``height'' profiles through the data are shown in [Fig.~\ref{AFM_AtRes}(c)]. These data unambiguously demonstrate the pure atomically resolved picture of graphene on Ir(111) obtained in NC-AFM in attractive regime: in AFM images as well as in the corresponding profile of the $z$-channel the distance between two carbon atoms is $1.5\pm0.1$\,\AA. The difference in the hight for two different atoms in the unit cell can be explained by the fact that for the ATOP place two carbon atoms occupy different high symmetry atomic positions with respect to the underlying Ir(111) substrate. Also the slight variation in the graphene unit cell of the interaction strength between carbon atoms and Ir substrate might lead to the different potential variation influencing the imaging contrast on the atomic scale. From the other side, as expected, for the tunnelling current we observe only one single maxima in the centre of the carbon ring, where mostly tunnelling from the electronic states of the underlying Ir(111) substrate takes the place at the ATOP positions. 

\subsection{Constant height imaging}\label{CHAFMSTM}

Previously discussed in Sec.~\ref{STMandBIAS} and Sec.~\ref{CFSAFM} CC STM and CFS AFM results indicate the moir\'e cell selective interaction strength between a graphene layer on Ir(111) and the apex of the scanning sensor. These facts tell us that in case of adsorption of guest atoms on top of this system, they will nucleate at the particular sites of the moir\'e structure. In order to get an additional information about energetics of this adsorption processes the more sophisticated experiments are required. Next sections will be devoted to the discussion of the results of such experiments, which allow to go deeply inside of these effects.

\begin{figure}
\center
\includegraphics[scale=0.5]{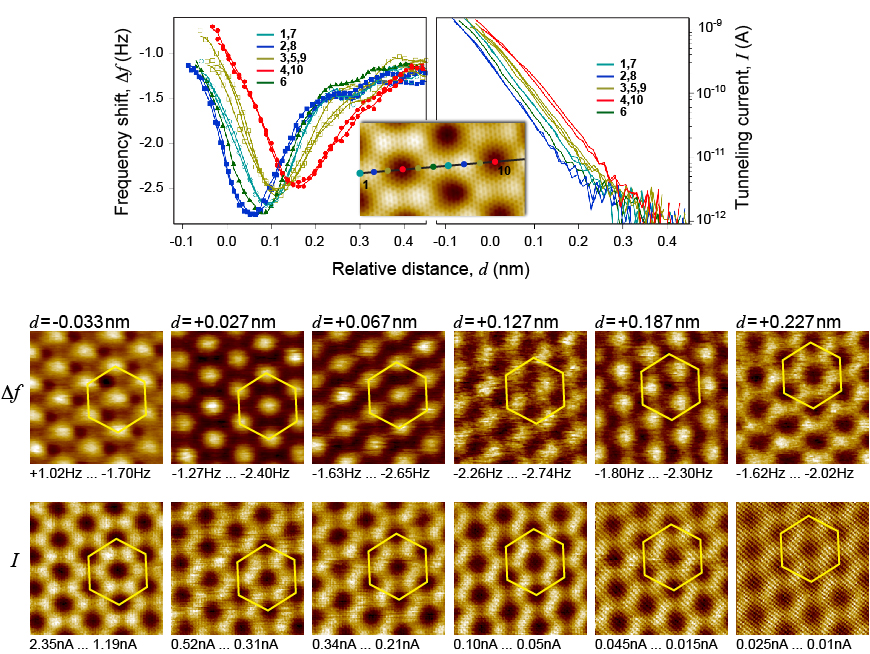}
\caption{Frequency shift ($\Delta f$) and tunnelling current as a function of the distance between tip and sample are shown on the left and right panels, respectively, of the upper part. The inset shows the corresponding STM image (scanning parameters: scan size is $10.5\times 10.5$\,nm$^2$, $U_T=+0.05$\,V and $I_T=0.4$\,nA, $A=100$\,pm and the path for the $z$-spectroscopy. Lower part shows the series of the CH AFM images [$\Delta f(x,y)$ and $I_T(x,y)$ maps] acquired at different $z$-positions of the scanning tip. The maximum (bright) and minimum (dark) values of $\Delta f$ and $I_T$ are marked under every image. The colour scale is linear.}
\label{AFM_CH}
\end{figure}

Fig.~\ref{AFM_CH} (upper part) shows a series of $\Delta f(z)$ and $I_T(z)$ curves measured at several points (1-10) along the line shown in the STM image of graphene/Ir(111) which is presented as an inset. Here the feedback set-point (constant current) from STM measurements was taken as a reference ``0'' point for the $z$-coordinate and then it was used for the correction of the $z$-scale. Generally, following the sequence of equations for the description of interaction between scanning tip and the sample surface,

\begin{equation}
F_z(z) = -\frac{\partial E(z)}{\partial z}, \Delta f(z) = -\frac{f_0}{2k_0}\frac{\partial F_z(z)}{\partial z},
\end{equation}
\noindent
where $E(z)$ and $F_z(z)$ are the interaction energy and the vertical force between tip and sample, respectively, $f_0$ and $k_0$ are the resonance frequency and the spring constant of the sensor, correspondingly, one can separate attractive and repulsive parts in the interaction. For the large distances between tip and sample (more than $5$\,\AA) the van der Waals as well as electrostatic attractive forces determine the interaction. As soon as the distance is decreased the attractive and repulsive chemical forces start to play a dominant role and here the chemical as well as atomic contrast can be routinely imaged~\cite{Seo:2007fu,Sugimoto:2007jo,Albers:2009ig,Sugimoto:2010gb,Baykara:2010a,Boneschanscher:2012bg,Setvin:2012de}.

Analysing the results presented in Fig.~\ref{AFM_CH} (upper panel) one can immediately conclude that there is a clear difference between $\Delta f(z)$ curves measured at different places of the graphene moir\'e structure on Ir(111) indicating the different interaction strength between tip and the sample. Considering the two limited cases, ATOP and FCC places of graphene/Ir(111), we can see that curves are shifted with respected to each other not only in $z$-direction, but also the minimum value of $\Delta f$ is different for all high-symmetry places. The difference in $z$-direction is $0.097$\,nm which is by $0.03$\,nm large compared to the value of graphene corrugation extracted from the corresponding STM image [inset of Fig.~\ref{AFM_CH} (upper panel)]. From these facts we can conclude that there is a discrimination in the interaction strength between scanning oscillating tip and the high symmetry positions of the graphene/Ir(111) structure.

Considering the results of the $\Delta f(z)$-spectroscopy one can indicate the existence of the crossing points for different curves as a function of coordinate $z$. These interesting behaviour will define the imaging characteristics of the graphene/Ir(111) structure in the CH AFM mode as demonstrated here. The results of the CH AFM imaging are presented in Fig.~\ref{AFM_CH} (lower panel), where the sequence of $\Delta f(x,y)$ and $I(x,y)$ images for different relative distances $d$ (from the upper panel) between tip and sample are shown. The bright hexagon traces the corresponding unit cell of the moir\'e structure of graphene/Ir(111). One can clearly see that the observed imaging contrast in the $\Delta f$ channel strongly depends on the relative distance between tip and sample and for two limited positions $d_i=-0.033$\,nm and $d_f=+0.227$\,nm the contrast is fully inverted: bright places become dark and vice versa. At the same time, the imaging contrast for the $I(x,y)$ channel is always the same -- \textit{inverted} imaging contrast as was previously observed in the CC STM imaging of graphene/Ir(111) [Fig.~\ref{STM_LargeScale}(a)]. [Here we would like to point out that $\Delta f(x,y)$ and $I(x,y)$ channels in CH AFM experiments are acquired simultaneously allowing the careful tracking of all changes in the imaging contrast.]

The imaging contrast inversion in CH AFM for the $\Delta f$ channel can be easily understood on the basis of results presented in Fig.~\ref{AFM_CH} (upper panel): there are several crossing points for the $\Delta f(z)$-curves measured at different high symmetry positions of graphene/Ir(111) and the imaging characteristics will depend on the sequence of the $\Delta f$ curves at the particular value of the relative distance $d$. Considering two limit distances $d_i$ and $d_f$ we can see that for the first position the value of $\Delta f$ is more negative for the FCC positions and these places are imaged as dark in this case. In the second case the situation is opposite and the FCC positions are imaged as bright places. In the similar way the imaging contrast can be traced for other positions. The possible inaccuracy in the interpretation for the intermediate distances can be due to the possible thermal drift during scanning.

The results for the $I(x,y)$ channel shown in Fig.~\ref{AFM_CH} (lower panel) provide also very important information about STM imaging of the graphene/Ir(111) system. According to the results presented in this figure the variation of the hight between two limit distances $d_i$ and $d_f$ is $0.260$\,nm. According to the recently simulated CC STM images of graphene/Ir(111)~\cite{Voloshina:2013dq} the change of the distance between tip and graphene/Ir(111) is $0.25$\,nm when bias voltage is varied in order to fulfil the constant current conditions of the STM experiment and simulate the experimentally observed contrast inversion in CC STM. Following these two facts, we can conclude that the electronic contribution is prevailing the topographic one in the imaging contrast in CC STM and CH STM modes for the graphene/Ir(111) system. Here the imaging contrast can be explained by the existence of several hybrid states in the valence band of the system (see discussion above).

\begin{figure}
\center
\includegraphics[scale=0.55]{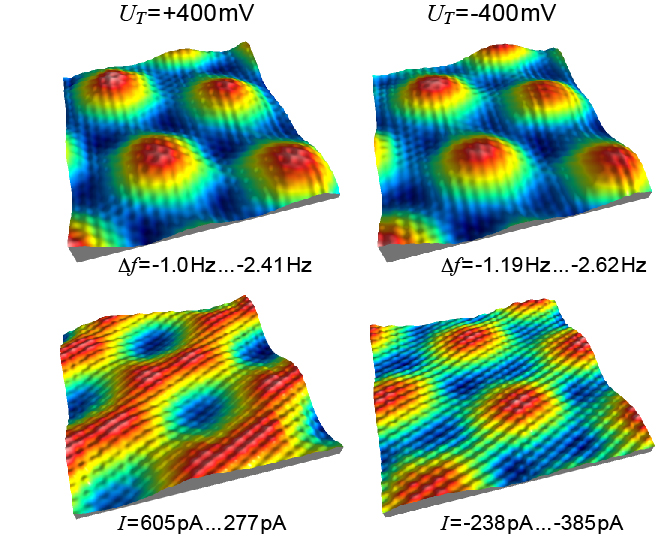}
\caption{Atomically-resolved CH AFM images of graphene/I(111) obtained at opposite bias voltages $U_T=\pm0.4$\,V; $\Delta f$ and $I$ channels are collected simultaneously. Imaging parameters: scan size $4.3\times 4.3$\,nm$^2$, $A=100$\,pm.}
\label{AFM_CH_diff-bias}
\end{figure}

CH AFM imaging of graphene/Ir(111) is not influenced by the changing of the polarity of the applied bias voltage. Fig.~\ref{AFM_CH_diff-bias} shows two atomically-resolved data sets for the $\Delta f(x,y)$ and $I(x,y)$ channels acquired at two opposite bias voltages $U_T=+0.4$\,V and $U_T=-0.4$\,V, respectively. Here two data sets ($\Delta f$ and $I$ are collected simultaneously) were obtained in consecutive order. One can see that the imaging contrast for the $\Delta f$ channel is the same demonstrating the absence of any ``cross-talk'' between two imaging channels (the slight variation of the $\Delta f$ values is due to the thermal drift). The first impression from the ``inversion'' of the contrast for the $I$-channel is simply due to the change of the sign of the bias voltage. Hence, two $I(x,y)$ images in the lower row of Fig.~\ref{AFM_CH_diff-bias} represent the similar imaging contrast. The difference in the absolute values of the tunnelling current is due to the different DOS above and below $E_F$ for the graphene/Ir(111) system.

\subsection{3D NC-AFM and STM}

The so-called 3D force and current microscopy and spectroscopy techniques~\cite{Albers:2009ig,Baykara:2010a,Pawlak:2012bx,Sugimoto:2012dc} are very powerful tools for the investigation of the electronic and structural properties of surfaces and interfaces. Generally, these methods allow to obtain three-dimensional data sets for the frequency shift of the oscillating scanning tip [$\Delta f(x,y,z)$] and the tunnelling current [$I(x,y,z)$], which then can be used for the calculation of the force map [$F(x,y,z)$] as well as the interaction energy landscape [$E(x,y,z)$] in the system.

\begin{figure}
\center
\includegraphics[scale=0.45]{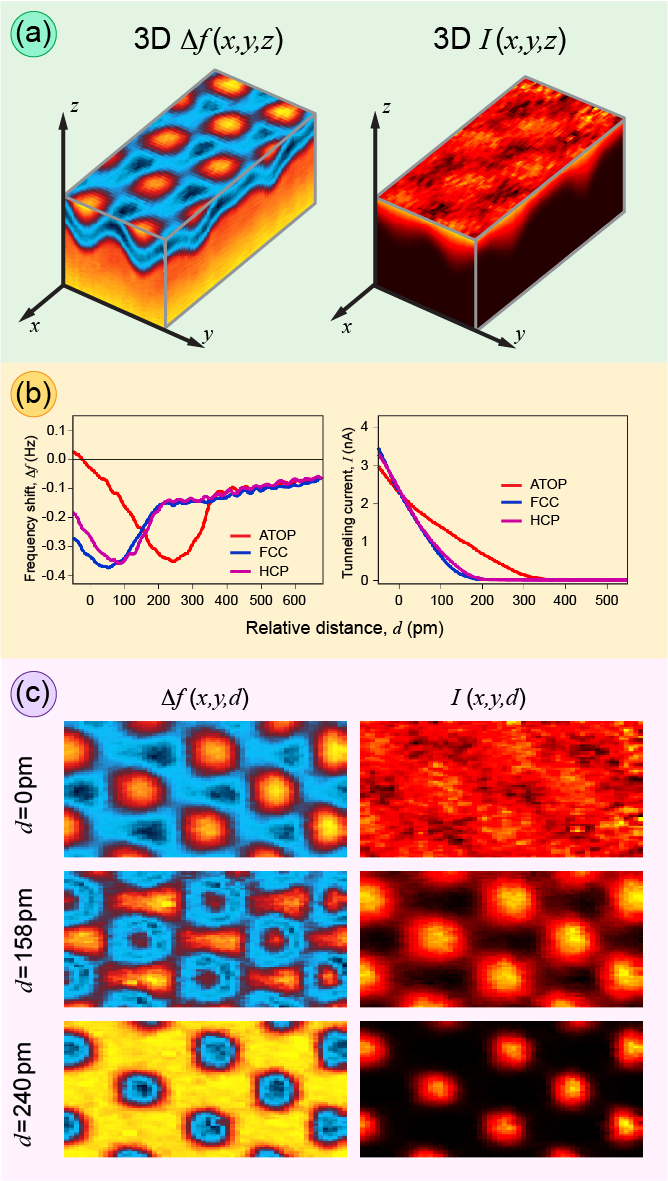}
\caption{(a) Full $\Delta f(x,y,z)$ and $I(x,y,z)$ (absolute value) 3D data sets and (b) the corresponding single $\Delta f(z)$ and $I(z)$ spectra extracted for the main high symmetry positions of the graphene/Ir(111) system. (c) Several constant height cuts through the 3D data sets from (a) taken at fixed $z$ coordinates. Measurement parameters: $4.4 \times 9.9$\,nm$^2$, $U_T=-0.1$\,V, $I_T=2.5$\,nA $A=300$\,pm. As supplementary material, we also present two movies that show the $z$-coordinate scan through the full $\Delta f(x,y,z)$ and $I(x,y,z)$ 3D data sets (available on request).}
\label{AFM_3D}
\end{figure}

Two approaches are usually used. In the first case the $\Delta f(x,y,z)$ and $I(x,y,z)$ data sets are collected on the $(x,y)$ grid in the short range of the $z$-coordinate, where the frequency shift and the tunnelling current are site sensitive. The long range $z$-coordinate scans are performed in some intermediate points in order to account for the possible thermal drift and for the estimation of the long range van der Waals and electrostatic interactions. The disadvantage of this method is that it is usually performed on the not very dense $(x,y)$ grid (due to the relatively long acquisition time) limiting the spacial resolution of the experiment. In the second approach a stack of the constant height $\Delta f(x,y)$ and $I(x,y)$ images is collected at different fixed $z$-positions of the scanning tip. Then these images are combined in one three-dimensional set of data, which is used in further analysis. Here two-dimensional images can be collected with as dense point grid as usually used in STM or AFM experiments. However, the possible drift might lead to the lost of some data points in the complete data set. Further details of these methods as well as all \textit{pro et contra} arguments for different approaches can be found elsewhere~\cite{Albers:2009ig,Sugimoto:2012dc,Sugimoto:2008bs,Heyde:2006el,Abe:2007jm}.

\begin{figure}
\center
\includegraphics[scale=0.55]{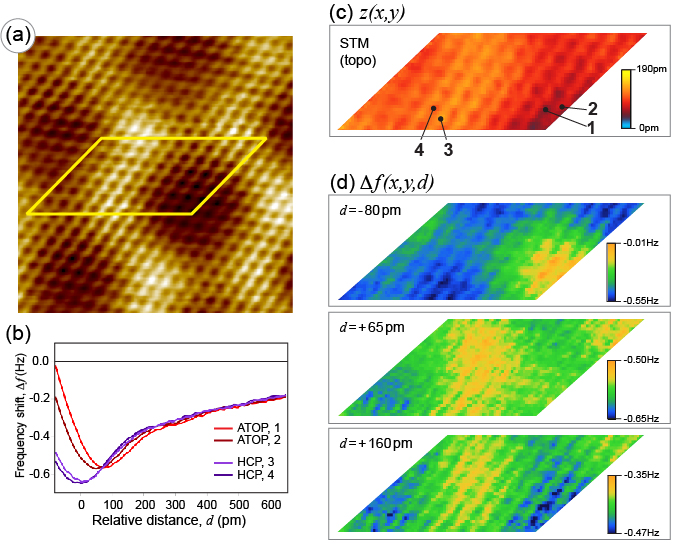}
\caption{(a) STM image of graphene/Ir(111) with the area marked by the yellow quadrangle where 3D full $\Delta f(x,y,z)$ and $I(x,y,z)$ 3D data sets were collected. (b) Representative $\Delta f(z)$ curves for the places marked in the topographic STM image shown in (c) by the corresponding digits. (d) Several constant height cuts through the 3D $\Delta f(x,y,z)$ data set taken at fixed $z$ coordinates. Measurement parameters: $U_T=-0.39$\,V, $I_T=1.6$\,nA, $A=300$\,pm. As supplementary material, we also present a movie that show the $z$-coordinate scan through the full $\Delta f(x,y,z)$ 3D data set (available on request).}
\label{AFM_3D_atres}
\end{figure}

In the present study the 3D data sets, $\Delta f(x,y,z)$ and $I(x,y,z)$, were collected on the grid, where single $\Delta f(z)$ and $I(z)$ curves were measured simultaneously (Figs.~\ref{AFM_3D} and \ref{AFM_3D_atres}). Here the feedback loop was stabilised in the CC STM mode allowing to follow the possible drift during scanning as well as for the calibration of the ``0'' off-set point for the $z$-spectroscopy measurements.

Fig.~\ref{AFM_3D} shows two 3D data sets, $\Delta f(x,y,z)$ and $I(x,y,z)$, measured for the moir\'e graphene structure on Ir(111). Presented data were corrected only for the drift in the $(x,y)$ plane (CC STM data were taken as a reference) without any additional smoothing or filtering of measured data. Compared to the data shown in Fig.~\ref{AFM_CH} the correction for the ``0''-point of the $z$-scale was not performed. However, the extracted from CC STM corrugation of graphene/Ir(111) reduce the distance between minima of the ATOP and FCC $\Delta f(z)$ curves only by $115$\,pm. The presented data give a possibility to get full information about local electronic structure and the local interaction strength in the different places of the graphene/Ir(111) structure and clearly demonstrate the effect of the inversion of the imaging contrast for the $\Delta f$ channel as a function of distance between tip and surface. Similar to the data presented earlier, the extracted single $\Delta f(z)$ curves for the high symmetry positions of graphene/Ir(111) have the so called crossing-points that explain the existence of the contrast inversion in CH AFM imaging. At the same time there is no contrast inversion in the $I$ imaging channel. The extracted \textit{horizontal} cuts from the 3D data sets are very similar to the data presented in Sec.~\ref{CHAFMSTM} and are CH AFM images of graphene/Ir(111), however collected on the less dense grid. Nevertheless, the presented data allow to follow the contrast inversion effect more carefully and to study the local interaction in the system more systematically.

The further step in the 3D AFM/STM studies of the graphene/Ir(111) was made upon decreasing of the probed area and increasing of the density of the scanning grid. The results of these experiments are shown in Fig.~\ref{AFM_3D_atres}. Here on the part of the surface of graphene/Ir(111) marked by yellow quadrangle in (a) a full 3D $\Delta f(x,y,z)$ and $I(x,y,z)$ data sets were recorded. Several representative examples of $\Delta f(z)$ curves extracted from the 3D data set are shown in (b) for the places marked by the corresponding numbers in the reference \textit{topographic} image obtained in CC STM mode shown in (c). This reference image was used (as in the previously discussed data set) to trace the possible drift during experiments as well as for the correction for the ``0'' reference point in the $z$-spectroscopy experiments. In the presented set all $\Delta f(x,y,z)$ data (b,d) were corrected for the corresponding ``0''-$z$-starting-point extracted from the \textit{topographic} CC STM data in (c). The panel (d) shows a sequence of the horizontal cuts of 3D $\Delta f(x,y,z)$ set taken at some $z$ positions taken from (b) and marked in the figure.

According to the results presented earlier the digitally marked places in the topographic STM image of graphene/Ir(111) [Fig.~\ref{AFM_3D_atres}(c)] correspond to ATOP (places marked by 1 and 2) and HCP (places marked by 3 and 4) high symmetry positions. Similar to the earlier discussed data here we can also see the inversion of the imaging contrast on the global scale for different places in the graphene moir\'e in CH AFM images extracted from the 3D $\Delta f$ data sets [see Fig.~\ref{AFM_3D_atres}(d)]. Increasing of the spatial resolution in such measurements immediately leads to the possibility to observe the atomically resolved contrast in the 3D $\Delta f$ data sets. The interesting observation here is that the inversion of the imaging contrast in the CH AFM data is observed not only on the global scale (moir\'e structure), but also on the atomic scale [Fig.~\ref{AFM_3D_atres}(d)]. For example, considering the two positions, marked by ``1'' and ``2'', of the the ATOP place we can see that the atomically resolved imaging contrast is fully reverted when one analyses two CH AFM images at $d=-80$\,pm and $d=+160$\,pm. The similar effect (but nor so pronounced) can be also seen for the HCP place (positions, marked by ``3'' and ``4''). The inversion of the imaging contrast is explained by the existence of the corresponding crossing points for the atomically resolved $\Delta f(z)$ curves similar to those observed during of the contrast inversion in the graphene moir\'e structure on Ir(111).

\subsection{KPFM imaging}

Kelvin probe force microscopy (KPFM) method is widely used for the investigation of the local variation of the work function of the material, which is extracted from the variation of the local contact potential difference ($U_{LCPD}$) between scanning metallic tip and the surface~\cite{Nonnenmacher:1991,Glatzel:2003fq,Sadewasser:2009jv,Melitz:2011}. In this method the NC-AFM measurements are performed and at the same time the DC ($U_{DC}$) and AC ($U_{AC}$) voltages are applied between sample and the conductive tip. Here the $U_{AC}$ voltage produces the oscillating electrostatic forces between tip and sample and the static $U_{DC}$ nullifies the oscillating electrostatic forces, which are result of the contact potential difference between tip and sample surface. A lock-in technique is used in order to measure the first harmonic of the signal, which is proportional to the difference of $U_{DC}$ and $U_{LCPD}$. In such a way the $U_{LCPD}$ value can be measured via applying the corresponding value of $U_{DC}$ in order to make the output of lock-in equal to zero. This method was successfully used for the imaging of the organic/metal interfaces~\cite{Palermo:2006jx}, graphene-based systems~\cite{Ziegler:2011ej,Wang:2011hi,Nazarov:2013hy,Druga:2013jz}, insulator/metal interfaces~\cite{Zerweck:2005kc,Cabailh:2012hm,Koch:2012by}, metallic clusters on the semiconducting surfaces~\cite{JingChung:2011iw}, defects of semiconducting surfaces~\cite{Rosenwaks:2004ke}, and other heterogeneous systems~\cite{Melitz:2011}. Recently the molecular and atomic resolutions were achieved~\cite{Zerweck:2007ks,Enevoldsen:2008jn,Nony:2009iv,Sadewasser:2009jv,Gross:2009jo,Mohn:2012gh} questioning the fundamental description of the work function of material and the electrostatic potential of atoms and molecules~\cite{Wandelt:1997ul}.

\begin{figure}
\center
\includegraphics[scale=0.65]{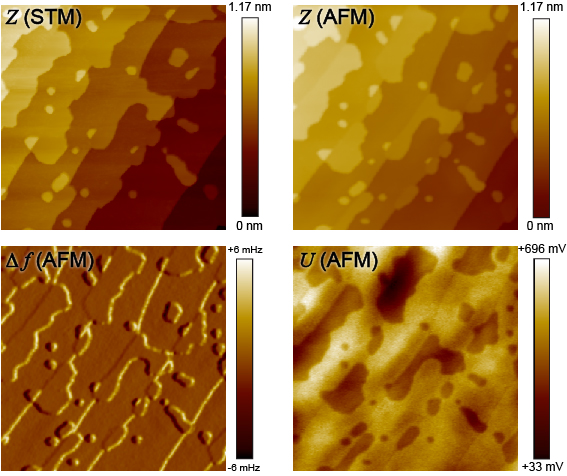}
\caption{Upper row presents two topographic images of the same area of Ir(111) covered by graphene flakes obtained in CC STM [$Z$ (STM)] and CFS AFM [$Z$ (AFM)] modes. Lower row shows $\Delta f$ and $U_{LCPD}$ channels which were recorded on the same area during NC-AFM imaging. Measurement parameters: $300 \times 300$\,nm$^2$, $\Delta f=-1.72$\,Hz, $A=500$\,pm, $U_{AC}=100$\,mV, $f_{AC}=45$\,Hz.}
\label{KPFM_largescale}
\end{figure} 

KPFM is obviously an ideal technique for identification of the graphene single and multilayers on top of metallic surfaces due to the relatively large difference in the work functions of metallic surfaces and graphene as well as due very high sensitivity of the method to the local variation of the electrostatic potential of as low as 0.1\,mV~\cite{Nonnenmacher:1991}. Presently, in most cases, for the identification of the graphene coverage on top of metallic surfaces, STM or AFM methods are widely used. However, sometimes these measurements can not give an unambiguous answer about precise location of graphene flakes. For example we can compare large scale STM and AFM topographic images of the same area presented in Fig.~\ref{KPFM_largescale} (upper row). Here we can see that in some cases the identification os the borders of the graphene flakes is not very easy. In some cases the use of the other parallel acquired channels ($I_T$ in STM or $\Delta f$ in AFM) in the analysis can help to be more precise. However, the only microscopic method here, which can give an unequivocal answer, is KPFM [Fig.~\ref{KPFM_largescale} (lower row)]. Graphene flakes can be identified as dark areas due to the lower work function of graphene compared to the one of Ir(111). The observed $U_{LCPD}$ contrast for the KPFM image of graphene/Ir(111) is 663\,mV. If we take this value as a maximum one for the reduction of the work function of Ir(111) upon graphene covering then we find that this value is smaller compared to the one of $1.1\pm0.3$\,eV and $1.6$\,eV extracted from STM $I(z)$ and thermionic emission experiments in low-energy electron microscopy (LEEM), respectively~\cite{Forster:2012aa,Starodub:2012cb}. The smaller value of $U_{LCPD}$ obtained from our KPFM measurements is due to the averaging effect of the macroscopic scanning tip with sharp apex that can lead to the reduction of the observed signal~\cite{Jacobs:1998vo,Baier:2012ey}.

\begin{figure}
\center
\includegraphics[scale=0.55]{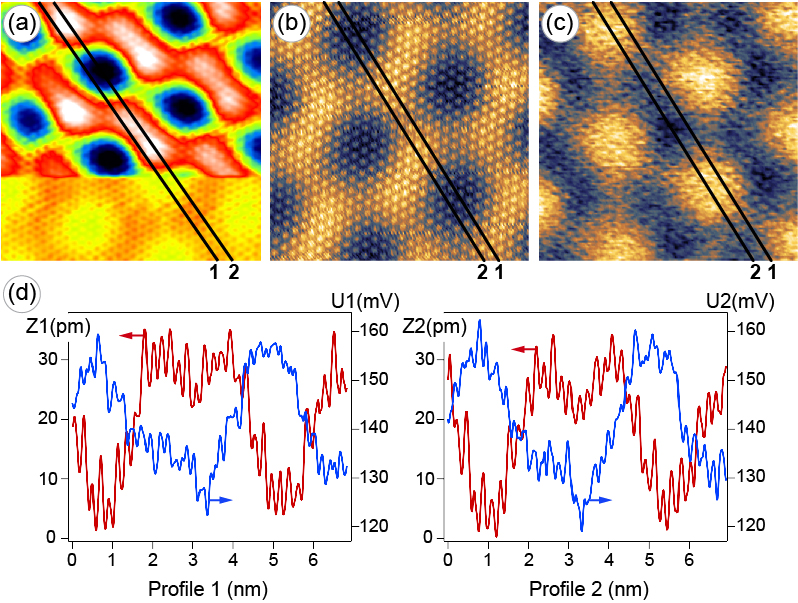}
\caption{(a) Combined STM/AFM measurements of graphene/Ir(111) where change between two modes was perfomed ``on-the-fly'' (CC STM -- top, CFS AFM -- bottom). Results of the KPFM experiment on the graphene moir\'e structure on Ir(111): (b) topography signal measured in CFS AFM mode and (c) simultaneously measured local contact potential difference. (d) Topography ($Z1$ and $Z2$) and local contact potential difference ($U1$ and $U2$) profiles extracted lines marked by ``1'' and ``2'' in (b) and (c), respectively. Measurement parameters: (a) $U_T=0.13$\,V, $I_T=1.8$\,nA, $\Delta f=-3.6$\,Hz, $A=300$\,pm, (b,c) $\Delta f=-3.6$\,Hz, $A=300$\,pm, $U_{AC}=100$\,mV, $f_{AC}=45$\,Hz.}
\label{KPFM_smallscale}
\end{figure}

The effect of the influence of the macroscopic tip on the imaging contrast in KPFM measurements can be reduced in the small scale measurements (Fig.~\ref{KPFM_smallscale}). Here the combined results of (a) STM/AFM and (b,c) AFM/KPFM studies of the graphene moir\'e structure on Ir(111) are presented. In order to carefully trace the imaging contrast in CFS AFM mode we perform ``switch-on-the-fly'' between STM and AFM [Fig.~\ref{KPFM_smallscale}(a)]. As was discussed earlier, the imaging contrast in AFM strongly depends on the $\Delta f$ set point and/or on the distance between oscillating scanning sensor and the sample surface. These results allow to carefully identify all high symmetry positions in the moir\'e structure and here the profile ``2'' on all images corresponds to the line which goes through the centres of two ATOP positions extracted from STM data [see Fig.~\ref{structure}(a)]. From the presented AFM/KPFM data [Fig.~\ref{KPFM_smallscale}(b,c)] one can clearly see the variation of local contact potential difference in the moir\'e unit cell of graphene/Ir(111). The variation of the $U_{LCPD}$ value is $\approx 35$\,mV with maximum and minimum values corresponding to ATOP and FCC positions of graphene/Ir(111), respectively. The corresponding topography and LCPD profiles taken through lines ``1'' (connecting extrema of the topographic image obtained in CFS AFM) and ``2'' (connecting extrema of the topographic image obtained in CC STM) are shown in [Fig.~\ref{KPFM_smallscale}(d)], respectively. From these profiles one can clearly see that the minimum value of the electrostatic potential corresponds to the FCC position with small potential barrier between FCC and HCP positions. This fact explains the preferential adsorption of atoms in FCC positions, supporting the conclusions, which were made earlier from AFM measurements. The similar result was recently published for the h-BN/Rh(111) system, where the $U_{LCPD}$ contrast of $\approx 60$\,mV in the moir\'e structure was measured by means of KPFM~\cite{Koch:2012by}.

It is interesting to note that the trace of the atomic resolution is clearly visible in the KPFM data presented in [Fig.~\ref{KPFM_smallscale}(c)] with $\Delta U_{LCPD}\approx 3-5$\,mV. In general, the observation of the atomic contrast in the $U_{LCPD}$ data can not be considered as a surprise as it was detected earlier in similar measurements on insulating and semiconducting surfaces~\cite{Bocquet:2008aa,Enevoldsen:2008jn,Nony:2009iv,Sadewasser:2009jv}. Previously it was discussed that many factors can determine the detection of the atomic resolution during KPFM measurements, like resistivity of the sample or tiny metallic tip, distance between tip and sample, etc.~\cite{Nony:2009iv,Weymouth:2012ct}. Therefore the presented results on the observation of the atomic contrast in the KPFM experiments on graphene/Ir(111) are still under internal discussion and will be the topic of the further publication. 

\subsection{Theoretical modelling}

\begin{figure}
\center
\includegraphics[scale=0.65]{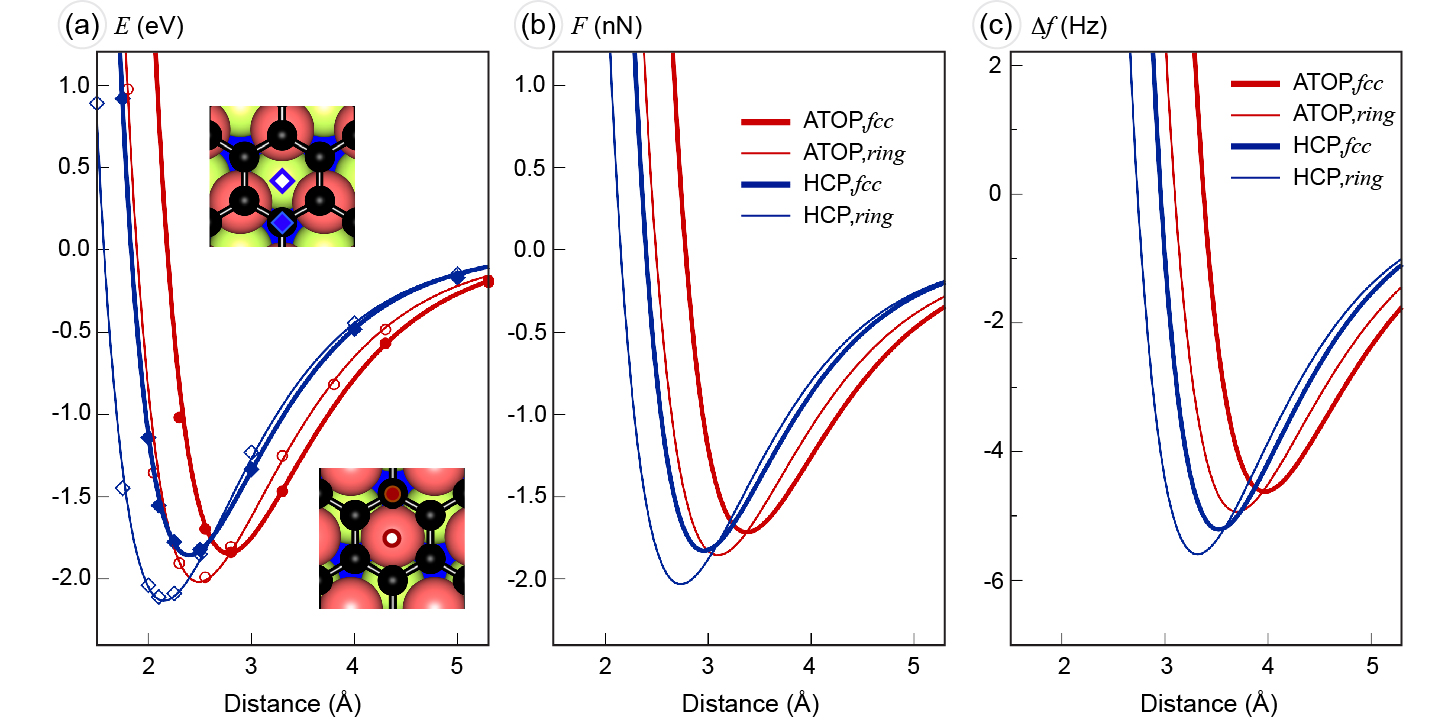}
\caption{(a) Calculated interaction energy, $E(z)$, between 5-W-atom pyramid and graphene/Ir(111) for 4 positions of the moir\'e unit cell (symbols of the calculated energy correspond to the symbols of the different places in the moir\'e structure shown in the inset). Data were fitted by the Morse potential and the corresponding curves are shown by thick and thin lines. (b,c) Corresponding force, $F(z)$, and frequency shift of the sensor, $\Delta f(z)$, extracted from (a) via differentiation of fitted curves for $E(z)$. }
\label{AFM_theory_curves}
\end{figure}

\textit{NC-AFM.} The results of the NC-AFM experiments were modelled via consideration of the interaction between tip and sample. Due to the large size of the graphene moir\'e unit cell on Ir(111), the apex of the scanning tip was modelled by the 5-W-atom pyramid~\cite{Voloshina:2013dq}. The interaction energy as a function of the distance, $E(z)$, between this model tip and the surface of graphene/Ir(111) was calculated for the ATOP and HCP high-symmetry positions in the moir\'e unit cell and correspondingly for two atomic places in every position, in the centre of the carbon ring (\textit{ring}) and above the carbon atom in the $fcc$ position (\textit{fcc}), i.\,e. above Ir(S-2) [inset of Fig.~\ref{AFM_theory_curves}(a)]. The results of these calculations are compiled in Fig.~\ref{AFM_theory_curves}, where (a) interaction energy between graphene/Ir(111) and 5-W-atom cluster, (b) corresponding calculated force, and (c) the frequency shift are presented for 4 different atomic places in the moir\'e structure. In all cases the interaction energy was calculated for several distances between tip and sample and then these results were fitted by the Morse potential [corresponding curves are shown by solid lines in Fig.~\ref{AFM_theory_curves}(a)] as the most suitable one for the description of the interaction between metal cluster and the graphene surface~\cite{Loske:2009}. Furthermore, obtained curves were used for the calculation of the force, $F(z)$, and the frequency shift, $\Delta f(z)$, curves via formulas $F(z)=-\partial E(z)/\partial z$ and $\Delta f(z)=-f_0/2k_0\cdot\partial F(z)/\partial z$, respectively.

The presented results clearly demonstrate the possibility to discriminate between different high-symmetry positions in the graphene moir\'e structure on Ir(111) as well as between different atomcaly-resolved places in the carbon rings. First of all, the value of the interaction energy is the same for the place above $fcc$ carbon atom in both ATOP and HCP positions due to the similar local chemical interaction in both cases and the relatively large C-Ir distance in both positions. At the same time there is a clear difference in the value of the maximal interaction strength for the position at the centre of the carbon ring for both ATOP and HCP position. This is due to the fact that at the ATOP position the Ir(S) atom is placed below the centre of the carbon ring giving the stronger repulsive interaction between W-cluster and the surface compared to the HCP position where Ir(S-1) atom is placed below the centre of the carbon ring. It is interesting to note that the extracted $F(z)$ and $\Delta f(z)$ curves [Fig.~\ref{AFM_theory_curves}(b,c)] show the rate and the curvature of the $E(z)$ curve demonstrating how fast and strong the attractive interaction is distributed above the graphene/Ir(111) surface. 

\begin{figure}
\center
\includegraphics[scale=0.55]{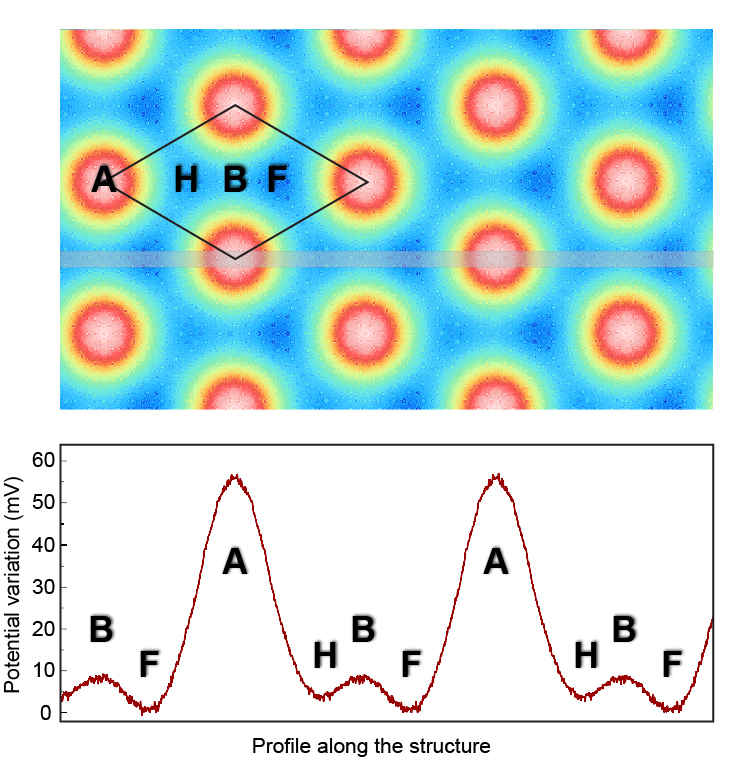}
\caption{2D map of the local electrostatic potential variation (upper panel) as well as the corresponding profile averaged over the grey rectangular (lower panel) on the surface of the graphene/Ir(111) system. The moir\'e unit cell is marked by the black rhombus in the upper panel. The capital letters mark the corresponding high-symmetry positions of graphene/Ir(111).}
\label{AFM_theory_KPFM}
\end{figure}

Fig.~\ref{AFM_theory_curves}(c) shows the calculated $\Delta f(z)$ curves for all above mentioned places in the moir\'e structure allowing the direct comparison with the experimental data. Firstly, the presented data qualitatively describe the observed effects shown in Fig.~\ref{AFM_3D_atres}: shape of curves and the respective sequence as well as the observed effects of the imaging contrast inversion in the moir\'e structure and on the atomic scale (there are several corresponding crossing points for the calculated $\Delta f(z)$ curves). There is a clear correspondence between following experimental and theoretical curves, respectively: ATOP,1 -- ATOP,$fcc$; ATOP,2 -- ATOP,$ring$; HCP,3 --HCP,$fcc$; HCP,4 -- HCP,$ring$. Surprisingly there is very good agreement for the differences in positions of the minima of the corresponding curves within one high-symmetry position. For the ATOP position theory gives $0.28$\,\AA\ vs. $0.26$\,\AA\ of the experimental value; for the HCP position theory gives $0.20$\,\AA\ vs. $0.11$\,\AA\ of the experimental value. At the same time, quantitatively, the calculated $\Delta f(z)$ curves are different by approximately factor of 7 from the experimentally obtained results.  There are several reasons that can explain this discrepancy: (i) the structure of the apex of the scanning tip is unknown in the experiment; (ii) the size of the W-cluster which models the scanning tip might be too small to mimic all interactions between tip (microscopic and macroscopic) and sample; (iii) during single $z$-point DFT calculations the system was not relaxed (however, the shape of experimental $\Delta f$ curves is reproduced); (iv) there is always some drift of the $z$-coordinate of the piezo scanner and of the resonance frequency of the oscillating sensor during time consuming 3D NC-AFM experiment that can lead to some uncertainty in the $z$ and $\Delta f$ values.\\
\textit{KPFM.} The optimised graphene/Ir(111) structure was used for the calculation of the distribution of the local work function in the system in order to compare with the results of the KPFM measurements. To extract the work function variation from DFT calculations, the local potential in the vacuum was determined by choosing the plane in the middle between the Ir slabs and then we subtracted the Fermi energy sampled over a fine surface mesh to account for all local C-Ir positions. Furthermore, the obtained potential distribution 2D map was smoothed with a two-dimensional Gaussian filter with a width of $3$\,\AA\ in order to take into account the STM tip convolution during KPFM experiments~\cite{Wang:2010ky}. The obtained result is presented in Fig.~\ref{AFM_theory_KPFM}, where the upper panel shows the calculated distribution of the local work function over the surface of the graphene/Ir(111) system with the corresponding profile (lower panel) averaged over the grey rectangular in the 2D map.

As seen from Fig.~\ref{AFM_theory_KPFM} the calculated variation of the work function for the graphene/Ir(111) system is in very good agreement with experimental data presented in Fig.~\ref{KPFM_smallscale}: our calculations completely reproduce the qualitative variation of the work function in the moir\'e cell -- it is higher for the ATOP place compared to other high-symmetry places. The calculated work function for the FCC high-symmetry place of graphene/Ir(111) is $4.366$\,eV, which we take as a reference value. The calculated total variation of $56$\,meV gives a local work function for the ATOP places of $4.422$\,eV, which is very close to the work function of the free-standing graphene ($4.48$\,eV)~\cite{Su:2007aa,Giovannetti:2008}. The experimentally observed KPFM contrast in the moir\'e cell of $\approx 35$\,meV discussed earlier is in reasonable agreement with theoretical value.

The calculated work function variation of $56$\,meV for graphene/Ir(111) can be compared with the previously calculated contrast for the graphene/Ru(0001) system: 500\,meV~\cite{Wang:2010jw} and $410$\,meV~\cite{Zhang:2011ky}. The experimentally determined value of the work function variation in graphene/Ru(0001) system is $220$\,meV~\cite{Wang:2010ky}. This strong variation in the later case can be attributed to the larger height variation of the carbon atoms in graphene/Ru(0001) and where regions with stronger bonding between graphene layer and Ru substrate are observed. This comparison immediately indicate a nearly free-standing character of the electronic states in a graphene layer on Ir(111). 

\section{Conclusions}

We employed scanning probe methods (constant current STM, constant frequency shift AFM, constant height STM/NC-AFM, and 3D STM/NC-AFM) for the investigation of the crystallographic and the electronic structure of the graphene moir\'e structure on Ir(111). These studies help us to understand the observed changes in the imaging contrast during STM and AFM measurements and are due to the variation of the local interaction strength between a graphene layer and the Ir(111) substrate. The variation of the local electrostatic potential was measured by means of the local Kelvin probe method during NC-AFM. These measurements indicate the different local electrostatic potential for different high-symmetry positions in the moir\'e structure of graphene/Ir(111). All experiments, STM and NC-AFM, were modelled in the framework of the DFT formalism with the van der Waals interaction taken into account. These calculations reproduce our experimental observations on the qualitative as well as on the quantitative level helping to understand the observed results. In the modelling of the NC-AFM results the scanning tip was modelled as a small cluster and the imaging contrast in the CH AFM mode was explained as a combination of the topographic and the local chemical contributions in the interaction strength between tip and surface. Also our DFT results reproduce quantitatively the observed variation of the KPFM signal in the moir\'e unit cell correctly identifying a local potential for all high-symmetry positions in the structure. The present studies demonstrate that the complementary studies by means of the different scanning probe methods accompanied by the state-of-the-art DFT calculations help to understand all structural and electronic contributions in the imaging of the carbon-based structures. This might be very important for the fabrication as well as for the characterisation of these systems.  

\section*{Acknowledgements}

E.V. acknowledges financial support from the DFG through the the Priority Program 1459 ``Graphene'' and computing facilities of the North-German Supercomputing Alliance (HLRN). Edoardo Fertitta (FU Berlin) is acknowledged for the technical assistance.

%
%
%

\section*{References}


\end{document}